\documentclass[twocolumn]{webofc} 
\usepackage[varg]{txfonts}   
\usepackage{graphicx,amsmath}
\begin{document}
\title{Two-phonon structures for beta-decay theory}
\author{A. P. Severyukhin\inst{1,2}\fnsep\thanks{\email{sever@theor.jinr.ru}}
\and N. N. Arsenyev\inst{1}\and I. N. Borzov\inst{3,1}\and R. G. Nazmitdinov\inst{4,1,2}
\and S. {\AA}berg\inst{5}}

\institute{
Bogoliubov Laboratory of Theoretical Physics, Joint
Institute for Nuclear Research, 141980 Dubna, Moscow region,
Russia \and Dubna State University, 141982 Dubna,
Moscow region, Russia \and National Research Centre ``Kurchatov Institute'',
123182 Moscow, Russia \and Departament de F\'{\i}sica,
Universitat de les Illes Balears, E-07122
Palma de Mallorca, Spain \and Mathematical Physics, Lund University,
PO Box 118, S-22100 Lund, Sweden}

\abstract{
The $\beta$-decay rates of $^{60}$Ca have been studied within a
microscopic model, which is based on the Skyrme interaction T45 to construct
single-particle and phonon spaces.
We observe a redistribution of the Gamow-Teller
strength due to the phonon-phonon coupling, considered in the model.
For $^{60}$Sc, the spin-parity of the
ground state is found to be $1^+$. We
predict  that the half-life of $^{60}$Ca is 0.3 ms,
while the total probability of the $\beta x n$
emission is 6.1\%.
Additionally, the random matrix theory has been applied to analyse
the statistical properties
of the $1^+$ spectrum populated in the $\beta$-decay
to elucidate the obtained results.
}

\maketitle

The multi-neutron emission is basically a multistep process
consisting of (a) the $\beta$-decay of the parent nucleus (N, Z)
which results in feeding the excited states of the daughter
nucleus (N - 1, Z + 1) followed by the (b) $\gamma$-deexcitation to
the ground state or (c) multi-neutron emissions to the ground
state of the final nucleus (N - 1 - X, Z + 1) [see, e.g.,
Ref.~\cite{b05}].
There is a strong need in a consistent estimate
of the multi-neutron emission
for the analysis of radioactive beam experiments and
for modeling of astrophysical r-process. Recent experiments
indicate on evidence of strong shell effects in exotic calcium
isotopes~\cite{w13,s13}. Therefore,
the $\beta$-decay
properties of neutron-rich isotope $^{52}$Ca provide valuable
information~\cite{h85}, with important tests of theoretical
calculations (see, e.g., Ref.~\cite{s17}). In the report 
we  demonstrate the importance
of the phonon-phonon coupling (PPC) on the
$\beta$-delayed multi-neutron emission of the neutron-rich
nucleus $^{60}$Ca discovered in~\cite{t18}.

One of the successful tools for studying charge-exchange nuclear
modes is the quasiparticle random phase approximation (QRPA) with
the self-consistent mean-field derived from a Skyrme
energy-density functional (EDF). Indeed, such QRPA calculations
enable one to describe the properties of the parent ground state
and Gamow-Teller (GT) transitions using the same EDF.
In the case of the $\beta$-decay of $^{60}$Ca, we use
the EDF T45 which takes into account the tensor force added with
refitting the parameters of the central interaction~\cite{TIJ}.
The set T45 gives enough positive value of the spin-isospin Landau
parameter calculated at the saturation density ($G_0'$=0.1),
and a reasonable description of the $Q_{\beta}$ and  $S_{xn}$ values
for the $\beta$xn-emission of the even neutron-rich Ca isotopes.
The pairing correlations are generated by a zero-range volume force
with a strength of -315 MeVfm$^{3}$, and a smooth cut-off at 10 MeV
above the Fermi energies~\cite{svg08}. This value of the pairing
strength has been fitted to reproduce the experimental neutron
pairing energy of $^{50,52,54}$Ca~\cite{s17,Severyukhin17c}.
The calculated $Q_{\beta}$-window of the $\beta$-decay of $^{60}$Ca and
$S_{xn}$ values  of $^{60}$Sc are shown in Fig.~1. There is the possibility
of the nonzero probability of one-, two-, three-, four- and
fifth-neutron emission. As expected, the largest contribution
to the calculated $\beta$-decay half-life comes from the $1_1^+$ state, which
structure is dominated by  one unperturbed configuration. The lowest
two-quasiparticle (2QP) state is the configuration $\{\pi1f_{7/2},\nu1f_{5/2}\}$.

We employ  the standard procedure~\cite{t05} to construct
the QRPA equations on the basis of HF-BCS quasiparticle states
of the parent nucleus.
The residual
interactions in the particle-hole channel and the particle-particle channel
are derived consistently from the Skyrme EDF. The eigenvalues of
the QRPA equations within the finite rank separable approximation are found
numerically as the roots of the secular equation for the cases of electric
excitations~\cite{gsv98,svg08} and charge-exchange
excitations~\cite{svg12,ss13}. It enables us to perform the QRPA calculations
in a large 2QP spaces. In particular, the cut-off of the
discretized continuous part of the single-particle spectra is
performed at the energy of 100~MeV. This is sufficient for exhausting
the Ikeda sum rule $S_{-}-S_{+}=3(N-Z)$.
\begin{figure*}[]
\includegraphics[width=1.8\columnwidth]{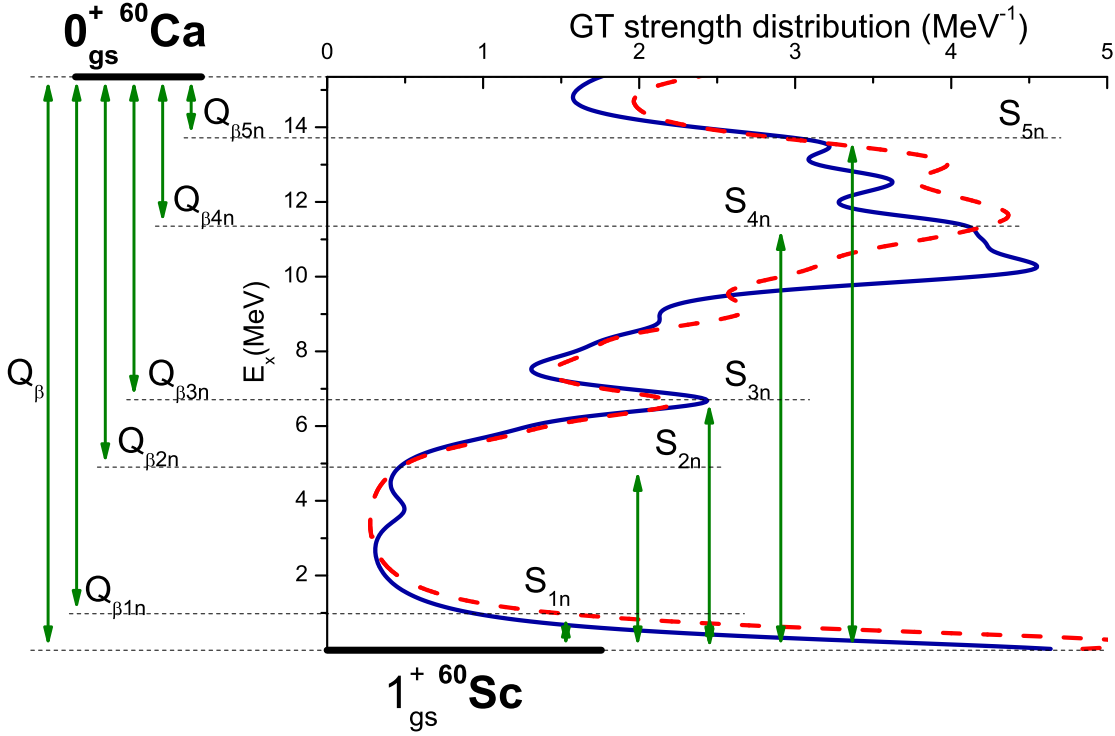}
\caption{GT strength distributions of $^{60}$Ca as functions
of the excitation energy of the daughter nuclei.
The calculations, taking into account the microscopic
and random coupling matrix elements between
the one- and two-phonon configurations, are
shown as solid lines and dashed lines, respectively.
The smoothing parameter 1~MeV is used for the strength distribution
described by the Lorentzian function.}
\end{figure*}

Taking into account the basic ideas of the quasiparticle-phonon model
(QPM)~\cite{Soloviev92,Kuzmin84}, the Hamiltonian is then diagonalized in a
space spanned by states composed of one and two QRPA
phonons~\cite{Severyukhin13,Severyukhin14},
\begin{eqnarray}
\Psi _\nu (J M) = \left(\sum_iR_i(J \nu )Q_{J M i}^{+}\right.
\nonumber\\
\left.+\sum_{\lambda _1i_1\lambda _2i_2}P_{\lambda _2i_2}^{\lambda
_1i_1}( J \nu )\left[ Q_{\lambda _1\mu _1i_1}^{+}\bar{Q}_{\lambda
_2\mu _2i_2}^{+}\right] _{J M }\right)|0\rangle~, \label{wf}
\end{eqnarray}
where $Q_{\lambda \mu i}^{+}\mid0\rangle$ are  the wave functions of the
neutron-proton one-phonon states having energy $\omega_{\lambda i}$ of
the daughter nucleus (N - 1, Z + 1);
$\bar{Q}_{\lambda\mu i}^{+} |0\rangle$ is the one-phonon excitation having energy
$\bar{\omega}_{\lambda i}$ of the parent nucleus (N, Z).
For the unknown amplitudes $R_i(J\nu)$ and
$P_{\lambda_2i_2}^{\lambda_1i_1}(J\nu)$ the variational principle
leads to the set of linear equations with the rank equal to the
number of one- and two-phonon configurations,
\begin{eqnarray}
(\omega_{\lambda i}-\Omega_\nu )R_i(J \nu ) +\sum_{\lambda _1i_1
\lambda_2i_2} U_{\lambda _2i_2}^{\lambda _1i_1}(J i)
P_{\lambda_2i_2}^{\lambda _1i_1}(J \nu )=0, \label{2pheq1}
\end{eqnarray}
\begin{eqnarray}
(\omega _{\lambda _1i_1}+\bar{\omega}_{\lambda _2i_2}-\Omega_\nu
)P_{\lambda _2i_2}^{\lambda _1i_1}(J \nu)\nonumber\\
+\sum\limits_i U_{\lambda _2i_2}^{\lambda _1i_1}(J i)R_i(J \nu
)=0. \label{2pheq2}
\end{eqnarray}
To solve these equations
 it is required to compute the Hamiltonian matrix
elements coupling  between
one- and two-phonon
configurations~\cite{Severyukhin13,Severyukhin14},
\begin{equation}
U_{\lambda _2i_2}^{\lambda _1i_1}(J i)= \langle 0| Q_{J i }
H \left[ Q_{\lambda _1i_1}^{+}\bar{Q}_{\lambda _2i_2}^{+}\right]
_{J} |0 \rangle.
\label{U}
\end{equation}
In order to construct the wave functions~(\ref{wf}) of the low-energy
$1^{+}$ states, in the present study we assume  that
the two-phonon configurations are
constructed from the $1^+$ and  $3^+$ charge-exchange phonons, and the $2^+$
and $4^+$ phonons. The redistribution of the GT strength due to the PPC
is mostly sensitive to the multi-neutron emission probability~\cite{Severyukhin17b}.

In the allowed GT approximation, the $\beta^{-}$-decay rate is
expressed by summing up the probabilities (in units of
$G_{A}^{2}/4\pi$) of the energetically allowed transitions
($E_{k}^{GT}{\leq}Q_{\beta}$) weighted with the integrated Fermi
function
\begin{equation}
  T_{1/2}^{-1}{=}
  D^{-1}\left(\frac{G_{A}}{G_{V}}\right)^{2}
  \sum\limits_{k}f_{0}(Z+1,A,E_{k}^{GT})B(GT)_{k},
\end{equation}
\begin{equation}
  E_{k}^{GT}{=}Q_{\beta}-E_{1^{+}_{k}},
\end{equation}
where $\lambda^{k}_{if}$ is the partial $\beta^{-}$-decay rate,
$G_{A}/G_{V}{=}1.25$ is the ratio of the weak axial-vector and vector coupling
constants and $D{=}6147$~s (see Ref.~\cite{Suhonen07}). $E_{1_{k}^{+}}$ denotes
the excitation energy of the daughter nucleus. As proposed in
Ref.~\cite{Engel99}, this energy can be estimated by the following
expression:
\begin{equation}
  E_{1^{+}_{k}}{\approx}\Omega_{k}-E_{\textrm{2QP},\textrm{lowest}},
\end{equation}
where $\Omega_{k}$ are the $1_{k}^{+}$ eigenvalues of
the equations~(\ref{2pheq1}), (\ref{2pheq2}), and
$E_{\textrm{2QP},\textrm{lowest}}$
corresponds the lowest 2QP energy, i.e.,
the energy $\{\pi1f_{7/2},\nu1f_{5/2}\}$ in the case of $^{60}$Ca.
Moreover, the ground state of $^{60}$Sc is found as $1^+$.
The wave functions allow us to determine GT transitions whose
operator is $\hat{O}_{-}{=}\sum\limits_{i,m}t_{-}(i)\sigma_{m(i)}$.
\begin{equation}
  B(GT)_{k}{=}\left|\langle{N-1},{Z+1};1_{k}^{+}
  |\hat{O}^{-}|N,Z;0_{gs}^{+}\rangle\right|^{2}.
\end{equation}
Since the tensor correlation effects are taken into account
within the $1p{-}1h$ and $2p{-}2h$ configurational spaces,
any quenching factors are redundant~\cite{Bertsch82}.

The difference in the characteristic time scales of the $\beta$-decay
and subsequent neutron emission processes justifies
an assumption of their statistical independence.
As proposed in Ref.~\cite{Pappas72}, the $P_{xn}$ probability of the
$\beta{xn}$ emission accompanying
the $\beta$ decay to the excited states in the daughter nucleus
can be expressed as
\begin{equation}
  P_{xn}{=}T_{1/2}D^{-1}\left(\frac{G_{A}}{G_{V}}\right)^{2}
  \sum\limits_{k^{\prime}}f_{0}(Z+1,A,E_{k^{\prime}}^{GT})
  B(GT)_{k^{\prime}},
\label{pxn}
\end{equation}
where the GT transition energy ($E_{k^{\prime}}^{GT}$)
is located within the neutron emission window
$Q_{\beta{xn}}{\equiv}Q_{\beta}-S_{xn}$. For $P_{1n}$ we have
$Q_{\beta{2n}}{\leq}E_{k^{\prime}}^{GT}{\leq}Q_{\beta{n}}$,
while for $P_{xn}$ this becomes
$Q_{\beta{xn}}{\leq}E_{k^{\prime}}^{GT}{\leq}Q_{\beta{x-1n}}$.
Since we neglect the $\gamma$-deexcitation of the daughter
nucleus, some overestimation of the resulting $P_{xn}$ values can
be obtained~\cite{Severyukhin17b}.

The largest contribution (93\%) in half-life comes from the $1_1^+$
state calculated with the PPC. The dominant contribution (94\%) in the
wave function of the first $1^+$ state comes from the $[1^+_1]_{QRPA}$
configuration which is mainly built on the configuration
$\{\pi1f_{7/2},\nu1f_{5/2}\}$. The inclusion of the PPC leads to a
redistribution of the GT strength and the fragmentation is shown
in Fig.~1. The excitation energies refer to the ground state of the daughter
nucleus $^{60}$Sc. The half-life $T_{1/2}$=0.3 ms and the total probability
of the $\beta x n$ emission of $P_{tot}$=4.8\% are calculated within the QRPA.
The inclusion of the two-phonon terms results in the same half-life and
$P_{tot}$=6.1\%. Using the large $\beta$-decay window, we obtain the unexpectedly
small value of $P_{tot}$, the effect which was predicted within the
one-phonon approximation before. Similar $P_{tot}$-value of 7.7\% was predicted
within the pn-RQRPA~\cite{Marketin16}, however with nearly 18 times longer
half-life of 5.3 ms. The DF3+cQRPA calculation predicted a 6 times
longer half-life of $T_{1/2}$=1.8 ms but also a low $P_{tot}$-value of 11.3\%~\cite{B18}.

A natural question arises: what the complexity of the configurational
space should be enough in order to obtain the half-life and the $\beta$-delayed
neutron emission at extreme $N/Z$ ratio? This restriction can be justified by
the rough estimate from the random matrix theory~\cite{Metha}. We generalize the
approach based on the ideas of the random matrix distribution of the coupling between
one-phonon and two-phonon states generated in the QRPA~\cite{saan17}. We find that the
distribution of the matrix elements~(\ref{U}) is well reproduced by
a truncated Cauchy distribution. The similar tendency is observed for the description
of a gross structure of the spreading widths of monopole, dipole, and quadrupole giant
resonances~\cite{saan17,saan18}. Considering truncated Cauchy distributions, according
to the central limit theorem, the resulting shape (the average of the sum) is driving
the Gaussian distribution. Since the distribution of the matrix elements~(\ref{U}) is
symmetric with a finite rms value, we may generate the random matrix elements
from the truncated Cauchy distribution or from a Gaussian distribution~\cite{saan18}.

With the motivation above, we assume that the matrix elements~(\ref{U}) can be replaced
by a random interaction where the matrix elements are Gaussian distributed random numbers,
\begin{equation}
P(U)=\frac{1}{\sigma\sqrt{2\pi}} \exp\left({\frac{-U^2}{2\sigma^2}}\right)\,,
\label{rand}
\end{equation}
with the rms value $\sigma$ calculated with the microscopic PPC. Solutions are ensemble
averaged over the random interaction and give the GT strength distribution, see Fig.~1.
The value $P_{tot}$=5.3\% and the GT strength distribution calculated with the random
coupling matrix elements are close to those that were calculated within the microscopic PPC.
We conclude that the present approach makes it possible to perform the new analysis of
the rates of the $\beta$-delayed multi-neutron emission. The vitality of the obtained
results enables to us to extend the validity of our approach to the next level of
simplifications.  Namely, considering the microscopic one-phonon states coupled
randomly to the two-phonons energies generated from the Gaussian Orthogonal Ensembles
distribution. The computational developments that would allow us to conclude on this point
are under way.

In summary, by means of the Skyrme mean-field calculations and considering the coupling between
the phonons, we have studied
the $\beta$-decay properties of $^{60}$Ca. Using the Skyrme interaction T45 in
conjunction with the volume pairing interaction, the unexpectedly low probability
of the $\beta x n$ emission is obtained. To check this, the statistical properties
of the $1^+$ spectrum populated in the $\beta$-decay are analyzed. The restriction
of the two-phonon configurations can be justified by the rough estimate from the
random matrix theory, which demonstrate the unimportance of other
two-phonon composition  on the half-life and the neutron-emission  probability.

\begin{acknowledgement}
This work was partly supported by
Russian Foundation for Basic Research under Grant nos.
16-02-00228 and 16-52-150003.
\end{acknowledgement}

\end{document}